\documentclass[showpacs, prb,twocolumn,preprintnumbers , superscriptaddress, aps]{revtex4-2}
 
\usepackage{color}
\usepackage{amsmath,amssymb}
\usepackage{pifont}
\usepackage{amssymb}  % More symbols
\usepackage{bbold}
\usepackage{float}
\usepackage{subfloat}
\usepackage{amsmath,amssymb}
\usepackage{mathtools}
\usepackage{physics}
\usepackage{breqn}

\usepackage[caption=false]{subfig}
\usepackage{tikz}
\usepackage{makecell}
\usepackage{subfig}
\usepackage{pifont}   % Ding symbols
\usepackage{graphicx} % Include figure files
\usepackage{dcolumn}  % Align table columns on decimal point
\usepackage{bm}       % bold math
\usepackage{multirow} % Table functions
\usepackage{placeins}% For making floats not move around everywhere
\usepackage[colorlinks]{hyperref}
\usepackage{mathtools}

\newcommand{\vect}[1]{\mathbf{#1}}

\usepackage{breqn}
\usepackage{mathrsfs}

\usepackage{multirow}

\begin{document}

\title{High-accuracy evaluation of non-thermal magnetic states \\ beyond spin-wave theory: applications to higher-energy states}

	\author{Wesley Roberts}
%	\email{roberts.w@northeastern.edu}
	\affiliation{Department of Physics$,$ Northeastern University$,$ Boston MA 02115$,$ USA}
    %\affiliation{U.S. Army Combat Capabilities Development Command Army Research Laboratory$,$ Adelphi$,$ MD 20783 USA}
    
    \author{Michael Vogl}
		\affiliation{Physics Department$,$
		King Fahd University
		of Petroleum $\&$ Minerals$,$
		Dhahran 31261$,$ Saudi Arabia}
      \affiliation{Interdisciplinary Research Center (IRC) for Intelligent Secure Systems$,$ KFUPM$,$ Dhahran$,$ Saudi Arabia}

     \author{Roderich Moessner}
		\affiliation{Max Planck Institute for the Physics of Complex Systems$,$ N\"{o}thnitzer Str. 38$,$ 01187 Dresden$,$ Germany
}
  
	\author{Gregory A. Fiete}
%	\email{g.fiete@northeastern.edu}
	\affiliation{Department of Physics$,$ Northeastern University$,$ Boston MA 02115$,$ USA}
    \affiliation{Quantum Materials and Sensing Institute$, $ Northeastern University$, $ Burlington$,$ MA$,$ 01803$ $ USA}
	\affiliation{Department of Physics$,$ Massachusetts Institute of Technology$,$ Cambridge MA$,$ 02139 USA}

\begin{abstract}
    We present an approximation scheme based on selective Hilbert space truncation for characterizing non-thermal states of magnetic systems beyond spin-wave theory. We study applications to states that are inaccessible through linear spin-wave theory, such as multi-magnon states and higher-energy states. Our approach is based on the existence of an exact representation of spin operators in terms of finite-order polynomials of bosonic operators. It can be applied to systems with and without a magnetically ordered ground state. The approximation exactly diagonalizes the bosonic Hamiltonian restricted to particular boson occupation subspaces, improving the conventional linear spin-wave approach and exponentially reducing the computing time relative to exact diagonalization schemes. As a test case, we apply the approach to a prototypical one-dimensional model - an XXZ spin chain with an applied magnetic field and antisymmetric exchange coupling. Here the antisymmetric coupling introduces a continuous parameter to tune the system away from its exactly solvable limit. We find excellent agreement between numerically exact eigenstates and eigenvalues and those found via the approximation scheme. Our approach applies not just to higher lying states but also to boson bound states, which could make them more accessible to theoretical predictions for comparison with experiment.
\end{abstract}

\maketitle

\section{Introduction}

The study of magnetically ordered systems conventionally proceeds via an analysis of their low-lying excitations, known as magnons or spin-waves \cite{YUAN20221,Barman_2021, Kruglyak_2010,10.1063/1.5109132}. 
The magnon degrees of freedom are expressed using a Holstein-Primakoff (HP) \cite{HP} bosonic representation of the spins, expanded in a Taylor series around the classical magnetic ground state configuration of the system \cite{doi:10.1146/annurev-conmatphys-031620-104715, Altland}. However, being a truncated Taylor series, this representation is only approximate and accurately captures only a relatively small region of the system's full eigenspace. 
Importantly for what we will consider, we note that while this conventional method is reliable for exploring the low-lying energy eigenstates of magnetic systems, it cannot accurately capture higher-energy eigenstates. 

Moreover, such situations are experimentally realizable due to the relative robustness of typical magnetic insulators; higher-energy non-thermal magnetic states can be excited and probed via neutron scattering \cite{PhysRevB.67.054414,PhysRevB.108.064408,PhysRevB.2.1362,Bai2023-ss,PhysRevLett.124.197203}. Most critically, these measurements can be performed without destroying the underlying magnetic order. Additionally, as magnetic excitations are low-energy, typically meV, non-destructive techniques such as terahertz spectroscopy \cite{Wang2018-yy,PhysRevLett.127.267201,PhysRevLett.118.207204,PhysRevB.98.094425} can probe magnetic states outside the reach of the typical single-magnon theoretical picture. Such experimental capabilities invite theoretical approaches that are currently lacking. We aim to address this challenge and fill this gap here.

A reliable theoretical apparatus for studying phenomena at 
higher energies than accessible via a linear spin-wave approximation should, therefore, capture states lying beyond those available within the conventional magnon methods (which are only capable of describing multi-magnon states in a non-interacting limit). One approach in the literature that partially addresses this issue is called non-linear spin-wave theory \cite{PhysRevB.79.144416}. However, this approach has the drawback of only narrowly avoiding divergences due to subtle cancellations of the $1/S$ expansion. This observation seems to necessitate a $1/S$ expansion. Therefore, the technique does not consider the entire structure of spin operators (i.e., all orders in $S$), which might lead to artifacts such as seemingly broken symmetries\cite{PhysRevResearch.2.043243}. Moreover, it does not explicitly enforce a finite maximum magnon occupation number per site (real systems only have finite spin). This work proposes a theoretical approach to treating non-thermal higher-energy states of a spin Hamiltonian approximately avoids some of the issues of non-linear spinwave theory described above. 
 
Our approach is founded on an exact representation of spin operators in terms of bosonic operators, resulting in a largely forgotten extended (and exact) Holstein-Primakoff expansion \cite{PhysRev.87.568} that was recently rediscovered \cite{PhysRevResearch.2.043243}.  After representing an arbitrary spin Hamiltonian in exact bosonic form, we introduce an approximation scheme via truncation of the Hilbert space, segmenting it according to subspaces defined by boson occupation number. The approximation then proceeds by solving the Hamiltonian within a restricted subspace, for example, by exact diagonalization. One key benefit of such an approach is that higher-energy regions may be selected and solved approximately with significantly less numerical overhead than diagonalization of the full Hamiltonian would require. Details of this approximation scheme and arguments for its range of validity are discussed in what follows. We also note that because a thermal mixed state is a linear combination of all eigenstates weighted by a Boltzmann factor, we can approximate such states by approximating the eigenstates and eigenvalues for larger portions of the Hilbert space, but our focus is elsewhere here.

We begin our discussion in Sec.\ref{Sec:Expansion}, presenting the exact bosonic representations on which our approximation scheme is based \cite{PhysRevResearch.2.043243}. Sec.\ref{Sec:Approx} then details the approximation scheme, outlining the procedure for dividing the entire Hilbert space into the relevant subspaces and restricting the Hamiltonian to a set of subspaces. In Sec.\ref{Sec:Model}, we present a one-dimensional (1D) spin chain model to which we apply our approximation scheme. This model was selected because it is solvable via exact diagonalization for finite system sizes, so the approximate higher-energy states can be compared to the exact results to check the accuracy of the approximation scheme.  In Sec.\ref{Sec:mag_gnd_state} we discuss how the structure of the magnetic ground state differs from the linear spin wave theory result. Sec.\ref{Sec:Controlled} then argues that the approximation scheme is well-controlled, particularly concerning any parameter that can be varied to take the system from one in which particle number is a conserved quantity to systems where eigenstates have contributions from multiple occupation number sectors. For concreteness and illustration, these arguments are applied to the spin chain. Finally, Sec.\ref{Sec:Results} presents the results of the approximation scheme. In particular, we compare approximate states (via wavefunction overlap) and eigenvalues to their exact counterparts. Excellent agreement is found, and we argue that this observation can be expected to remain valid even as the system size is increased in the thermodynamic limit. We also discuss the numerical savings achieved via our Hilbert space truncation, demonstrating the practical value of the method. We complete our paper by discussing how our scheme can also characterize boson bound states. In particular, we show that the exact boson representation can identify phases of a model exhibiting bound states that are lower in energy than single-boson states. Applied to magnons, this approach opens the door to material proposals for systems with low-lying bound states, which are more experimentally accessible than the higher-energy multi-magnon states one might often expect in magnetically ordered systems, especially from the non-interacting, linear spin-wave picture.

\section{Exact boson transformation} \label{Sec:Expansion}
We begin by introducing two exact boson representations of a spin system. This step is done to provide consistency checks and to allow a choice in perspective.
\subsection{Finite polynomial resummation of the Holstein Primakoff square root}
It was shown in Ref.\cite{PhysRevResearch.2.043243} that a finite polynomial resummation of the bosonic operator square roots that appear in the Holstein-Primakoff representation of spin operators exists. This finding makes it possible to rewrite the spin lowering operator $S^- = a^{\dagger}\sqrt{2S -a^{\dagger}a}$ (where $S$ is the length of the spin, and $a^{\dagger}/a$  creates/annihilates a boson), for example, as a series with a finite number of bosonic terms. An exact bosonic Hamiltonian for any spin system - even those without a magnetically ordered ground state - can be obtained this way.

In this work, we restrict ourselves to spin-1/2 systems for simplicity. Although, it should be noted that all the following methods apply equally well for $S > 1/2$. The exact transformation for the spin raising operator in a spin-1/2 system is \cite{PhysRevResearch.2.043243,PhysRev.87.568}
\begin{equation} \label{eq:exactHP}
    S^{+} = a - a^{\dagger}a^2.
\end{equation}
This result should be contrasted with the $\mathcal{O}(1/S)$ truncation of a conventional Taylor series, which is given as
\begin{equation}
    S^{+} = a - \frac{1}{2}a^{\dagger}a^2.
\end{equation}

To better understand the relation between both forms of $S^+$, it is necessary to clarify the relationship between the spin and the boson occupation bases for the physical Hilbert space. The physical Hilbert space has $2^N$ degrees of freedom for a system of $N$ spins-1/2. However, naively, there is no bound on the Hilbert space of bosons - we can have any number at each lattice site. However, note that there is a relationship between the occupation number and the spin at each site:
\begin{equation}
    S^z = \frac{1}{2} - a^{\dagger}a.
\end{equation}
This equation means that along the defining $z$-axis for the bosons, the Fock states are also spin-$z$ eigenstates. Thus, occupation numbers higher than $2S$ for a spin-S system are unphysical.

Eq.\eqref{eq:exactHP} is exact in the sense that it does not couple states in the physical spin Hilbert space to unphysical states in the occupation representation - thus, any composite spin operator will likewise not couple these physical and unphysical spaces. For example, if one prepares a system in a physical eigenstate, it cannot transform into an unphysical state through the action of physical spin operators. This statement is not valid for the conventional Taylor series, highlighting that it is necessarily an approximation at any finite order. Interestly, while this observation was already made in 1952 \cite{PhysRev.87.568} it is not well-known in the literature.  One reason may be that it is difficult to naively use the expressions for both perturbation theory and mean field theories without running into issues \cite{PhysRev.87.568,10.21468/SciPostPhys.17.5.139}. One of the advantages of our present work is that it sidesteps these issues by making restrictions to physical subspaces in combination with quasi-degenerate perturbation theory.

\subsection{Alternative resummation using quasi-bosons}
Below, we give an alternative and straightforward version of resummation for the Holstein Primakoff square root, using quasi-bosons \cite{GARBACZEWSKI197865} rather than regular bosons. Quasi-bosons have a maximum occupation number built into their algebra rather than relying on a segmentation of the Hilbert space into physical and unphysical sectors. This property amounts to the algebra:
$$\alpha_i^{2S+1} = 0,$$
$$[\alpha_i, \alpha^{\dagger}_j] = (1 - \frac{2S+1}{2S!}\alpha^{\dagger 2S}_i \alpha^{2S}_i)\delta_{ij}.$$

For demonstration purposes, consider the case of spin $S = 1$. (The idea generalizes readily to higher spins.) In the $S=1$ case, one can add at most two bosons to a site so that $\alpha^3 = 0$. Aside from this, one has the same exact definitions of the spin operators in terms of quasi-bosons as for bosons. 

For example, Taylor expanding the square root for $S^+$ gives
\begin{equation}
    S^+ = \sqrt{2}(1 - \frac{1}{2}\alpha^{\dagger}\alpha)^{1/2}\alpha = \sqrt{2}\sum_{n=0}^{\infty} c_n (-\frac{1}{2}\alpha^{\dagger}\alpha)^n \alpha,
\end{equation}
where $c_n = \frac{(-1)^{n-1}(2n)!}{4^n (n!)(2n-1)}$ is the expansion coefficient. The key observation is that for $S = 1$, the quasi-bosons satisfy the relation $(\alpha^{\dagger}\alpha)^n \alpha = \alpha^{\dagger}\alpha^2$ for $n>0$. Separating the $n=0$ term, we have 
\begin{equation}
     S^+ = \sqrt{2}c_0 \alpha + \sqrt{2}\sum_{n=1}^{\infty} c_n (-\frac{1}{2})^n \alpha^{\dagger}\alpha^2,
\end{equation}
where $c_0 = 1$. The second term is almost the Taylor series for $\sqrt{1 - 1/2}$. We only need to add back in the $n=0$ term for this series so that
\begin{equation}
     S^+ = \sqrt{2} \alpha + \sqrt{2}\sum_{n=0}^{\infty} [ c_n (-\frac{1}{2})^n] \alpha^{\dagger}\alpha^2 - \sqrt{2} \alpha^{\dagger}\alpha^2,
\end{equation}
which gives
\begin{equation}
     S^+ = \sqrt{2} \alpha + (1 - \sqrt{2}) \alpha^{\dagger}\alpha^2.
\end{equation}
We can compare this to the expansion for $S = 1$ that employs real bosons (rather than quasi-bosons) \cite{PhysRevResearch.2.043243,K_nig_2021}, which gives 
$$S^+ = \sqrt{2} a + (1 - \sqrt{2}) a^{\dagger}a^2 + (\frac{1}{\sqrt{2}}-1)a^{\dagger 2} a^3.$$

The final extra term for true bosons restricts the spin operators so that physical bosonic states are never coupled to unphysical ones. In an approximation scheme like the one presented below, in which we restrict to a physical Hilbert space by default, these two expansions are equivalent, and their differences may be ignored.

\section{Approximation scheme for states beyond the linear spin-wave regime} \label{Sec:Approx}

Having defined an exact representation of spins in terms of bosonic operators, we now turn to the Hilbert space truncation scheme. This scheme allows us to approximate higher-energy magnon states—i.e., magnetic eigenstates beyond the linear spin-wave approximation—in a bosonic occupation representation, and importantly to account for magnon-magnon interactions. 

First, let us step back and recapitulate the conventional approach employed when studying magnons. Typically, an HP transformation is performed with a large-$S$ (or low-boson occupation) approximation, which, while formally justified for $S \to \infty$, is valid in the linear spin-wave regime down to $S = 1/2$. 
This property allows one to drop all but bilinear boson terms in a Hamiltonian that is bilinear in spins. 

The most common situation where this is used is for an expansion around the state that minimizes the classical ground state energy \cite{doi:10.1146/annurev-conmatphys-031620-104715, Altland}. First, one has to transform to a frame where the $z$-axis of all spins points along the local spin direction. The resulting quadratic Hamiltonian of HP bosons can then be diagonalized by going to Fourier space (and in the case of non-number conserving systems such as antiferromagnets, introducing new bosons via a Bogoliubov-de Gennes transformation). The basis in which the Hamiltonian is diagonal yields the excitations of the problem, which are called magnons. 

The approximation scheme we detail below captures these low-lying degrees of freedom but also allows us to characterize higher-energy (non-thermal) states. The method also allows for the characterization of magnon bound states
- an exciting and active area of physics that can greatly benefit from additional theoretical insights \cite{Fukuhara2013-yl,PhysRevLett.95.077201,PhysRevLett.100.027206,PhysRevLett.102.037203,PhysRevB.67.054414,PhysRevB.108.064408,PhysRevLett.127.267201,PhysRevLett.124.197203,Nishida2013-zc,PhysRevResearch.2.033024,Wang2018-yy,PhysRevLett.125.187201,PhysRevB.101.180401,Wulferding2020-jf,PhysRev.132.85,H_C_Fogedby1980-xe}.

\subsection{Truncated Hilbert space}
\label{sec:TruncHS}
The basic strategy our approximation employs is to diagonalize a truncated Hamiltonian. As a first approximation, we restrict to states with one or fewer bosons. That is, we find eigenstates of the Hamiltonian, $H$, in this truncated Hilbert space, i.e. solutions to
\begin{equation}
    H^{(1)} \ket{\psi^{(1)}} = E^{(1)}_k \ket{\psi^{(1)}},
\end{equation}
where $H^{(1)}$ is a matrix of only the $\leq1$ boson matrix elements of $H$.

To illustrate the structure, we present below the full matrix form of $H^{(1)}$ for a two-site model restricted to one or fewer bosons globally, which could be written as
\begin{equation}
   H^{(1)} = 
    \begin{pmatrix}
        \bra{0,0} H \ket{0,0} & \bra{0,0} H \ket{1,0} & \bra{0,0} H \ket{0,1}\\
        \bra{1,0} H \ket{0,0} & \bra{1,0} H \ket{1,0} & \bra{1,0} H \ket{0,1} \\
        \bra{0,1} H \ket{0,1} & \bra{0,1} H \ket{1,0} & \bra{0,1} H \ket{0,1}
    \end{pmatrix},
\end{equation}
where $\ket{0,0}$ is the boson vacuum and $\ket{n_1,n_2} = (a^{\dagger}_1)^{n_1} (a^{\dagger}_2)^{n_2}\ket{0,0}/\sqrt{n_1!n_2!}$ are occupied states with $n_1$ bosons at site 1 and $n_2$ bosons at site 2. Approximate states and energies are then found by diagonalizing $H^{(1)}$. 

Including more total occupation numbers in the truncated Hilbert space should improve the approximation because larger parts of the space are modeled more accurately, which is indeed what we find. 

Alternatively, one need not only restrict to the low occupation number subspace but also to sectors with high occupation number - for example, states missing only one boson, or ``single-hole" states (recall that or finite spin there is a maximum number of physically allowable bosons). This approach can access the system's highest-energy (non-thermal) states. In this case, we consider an analogous problem to $H^{(1)}$, but in the subspace of states such as $\ket{\tilde{i}} = a_i \ket{F}$, where $\ket{F}$ is the state in which all sites are filled, i.e., the physical state with maximal occupation. In the spin representation, this is the state in which all spins are flipped relative to the classical ground state.

\subsection{Expansion around other classical extrema}
\label{sec:otherExtrema}
In addition to including larger portions of the eigenspace to improve the approximation scheme's performance, we note another useful approximation based on the exact bosonic representation. Because it is possible to carry out an exact expansion around any classical reference state, one can carry it out around other extrema of the classical energy - not only the classical ground state. Indeed, classical ordered states may even exist when the quantum ground state is not ordered - for such a state to be of interest, it would need to be stable against the emission of particles.

We note that generally, after local rotations of spin operators that align their $z$-component with any classical extremal or saddle-point configuration, one finds a Hamiltonian that is simplified: when the resulting Hamiltonian is expressed in terms of HP bosons, one finds that all linear boson operators $a_i$ and $a^{\dagger}_i$  vanish \cite{doi:10.1146/annurev-conmatphys-031620-104715}. Thus, one may perform what is essentially a spin-wave analysis (dropping higher-order interactions). This step allows finding accurate approximations to states with energies close to the extremal classical energy - much like a typical spin-wave analysis. This idea of expanding around an energy extremum is similar to a stationary phase approximation, where linear terms also vanish near an extremum, and the function is approximately Gaussian near its stationary points \cite{Bender}. The result is an exact Hamiltonian (before dropping any interaction terms) describing excitations in a harmonic approximation around a more highly excited configuration.

Excitations defined for different configurations are distinct - bosons around the classical maximum are not the same as those defined by the classical minimum. By defining our bosons for more general spin configurations that extremize the classical energy, we arrive at a bosonic Hamiltonian for which occupation corresponds to lowering the spin along a different quantization axis than the classical ground state.

We note that this also means that we can, for example, keep pairing terms and perform a standard spin-wave analysis in terms of high-energy bosons existing around these higher-energy configurations in a Bogoliubov-de Gennes formalism \cite{10.1093/ptep/ptaa151,doi:10.1146/annurev-conmatphys-031620-104715, PhysRevB.87.174427,PhysRevB.105.L100402}. We can, in principle, find bands,  
quantum geometric quantities (quantum metric and Berry curvature), topological properties, and so on for these highly excited states.

For a concrete example of extremal or saddle-point configurations away from the classical minimum and maximum, we consider the one-dimensional (1D) Ising model in an external field, $h$:
\begin{equation} \label{Eq:Ising}
    H = -J \sum_i S^z_i S^z_{i+1} -h \sum_i S^z_i,
\end{equation}
where $i$ indexes the lattice sites.

\begin{figure}[htp]
\centering
    \includegraphics[width=8cm]{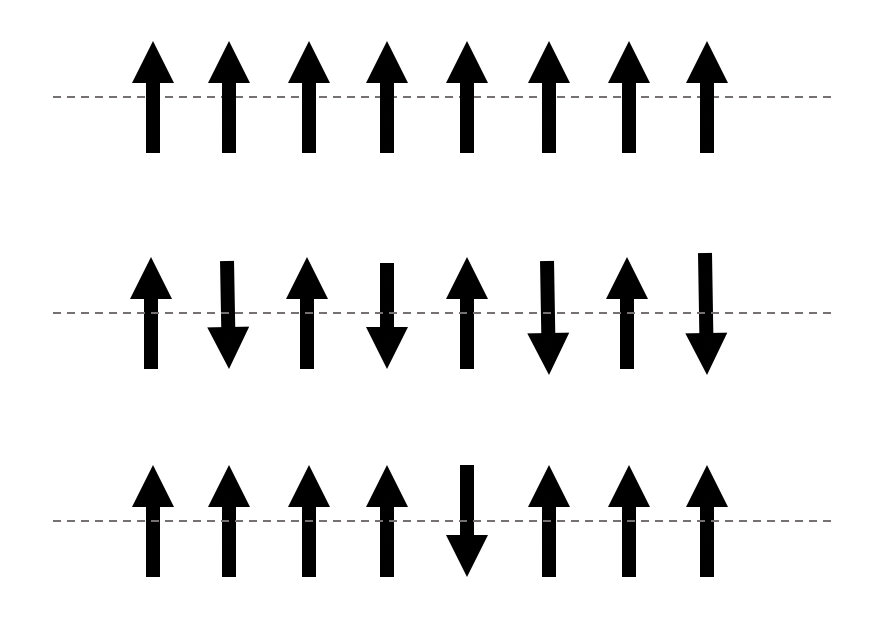}
   \caption{\label{Fig:IsingMetastable}Extremal configurations for the Ising chain in an applied field, Eq.\eqref{Eq:Ising}. The top configuration is the classically minimal energy ferromagnetic state, and the middle configuration is the classically maximal energy antiferromagnetic state. The bottom shows another classical configuration that could be expanded around - a saddle point.} 
    \end{figure}

To make the analysis as straightforward as possible, we consider cases in which $h < 2JS$ so that the dominant energy scale is the Ising coupling $J$. Letting $\theta_j$ be the angle of the $j$th spin from the $z$-axis, we can write the classical energy as
$$E = -JS^2 \sum_i \cos{\theta_i} \cos{\theta_{i+1}} - hS \sum_i \cos{\theta_i},$$
and we find that $\partial E/ \partial \theta_j = 0$ for all $j$ when
\begin{equation} \label{eq:extremize}
    JS^2 \sin{\theta_j}(\cos{\theta_{j+1}} + \cos{\theta_{j-1}}) + hS\sin{\theta_j} = 0.
\end{equation}

Condition \eqref{eq:extremize} clearly holds when $\theta_j = 0$ or $\pi$ at each site. The global minimum occurs when $\theta_j = 0$ everywhere, and the maximum when $\theta_j$ alternates between $0$ and $\pi$ - the antiferromagnetic state (of course, there are two degenerate global maxima). For all $j$, $\theta_j = \pi$  is a metastable point, and all other configurations of angles $0$ and $\pi$ are saddle points, as shown in Fig.\ref{Fig:IsingMetastable}. There is even an extremum where all spins have $\cos{\theta_j} = -h/2JS$. 

As one can see, even a simple model admits many classical extrema and saddle points: each spin needs to be aligned with the axis, but not necessarily the direction, of its exchange field. While such a situation looks quite unstable generically, spin-wave treatments in specific settings can nonetheless be surprisingly useful, see e.g.~Ref.\cite{amcr_dwd}. Each of these configurations can be used to define a bosonic representation in which linear terms disappear. Expansions around such a classical state permits one to accurately capture states with energies nearby - by truncating to a bosonic Bogoliubov Hamiltonian (if the aim is a non-interacting description to classify topological states or computing geometric responses) or by a truncation to different quasi-particle sectors (if one seeks an understanding of quasi-particle bound states in an energy sector).  

\section{1D model} \label{Sec:Model}
To test our approximation method, we apply it to a $S=1/2$ one-dimensional system, Eq.\eqref{eq:H_1D}, that can be solved numerically via exact diagonalization \cite{10.21468/SciPostPhys.2.1.003}. This approach allows us to compare approximate eigenvalues and eigenstates with the numerically exact solution. 

We consider an XXZ model with an applied magnetic field $h$ and Dzyaloshinskii–Moriya interaction (DMI) - also known as antisymmetric exchange - as our test case. Here $\vect{D}_{ij}$ is the antisymmetric coupling between sites $i$ and $j$. 

The DMI is crucial because it leads to the canting of spins in the magnetic ground state.
This term, in turn, produces non-number-conserving terms in the bosonic Hamiltonian - it means that eigenstates and eigenvalues in the single-particle Hilbert space will not be exact, as they would be for a number-conserving Hamiltonian. We choose a model with tunable DMI rather than, for example, a Heisenberg antiferromagnet because the DMI strength introduces a parameter by which we can tune the size of the pairing (number non-conserving) terms relative to the number-conserving ones. In contrast, there would be no such parameter for an antiferromagnetic model. 
The model is also desirable because, for a general choice of parameters, it is non-integrable \cite{JAHANGIRI202373} and therefore our results do not rely on any special symmetries of the problem.
The model Hamiltonian we consider is
\begin{dmath}
    H = -\sum_i \frac{J}{2} (S^{+}_i S^{-}_{i+1} + S^{-}_i S^{+}_{i+1}) -\sum_i\Delta J S^z_iS^z_{i+1} + \sum_i h S^z_i  + \sum_{<ij>}\vect{D}_{ij}\cdot (\vect{S}_i\cross \vect{S}_j),
    \label{eq:H_1D}
\end{dmath}
where we take a DMI vector configuration that alternates along the $x$-direction; $J$ is the nearest-neighbor exchange coupling, and $\Delta$ an anisotropy parameter that results in a coupling along the $z$-axis that differs from the ones for $x$- and $y$-axes.

We begin by rotating into local coordinate axes in which spins are given by $\vect{S}^l_i$
\begin{equation}
    \vect{S}_i = R_i \vect{S}^l_i.
\end{equation}
Here, the rotation matrix $R_i$ is chosen so that the $z$-axis of each spin aligns with the ground-state spin configuration.

We then perform an exact boson transformation on the spin-1/2 system, arriving at the following boson Hamiltonian (keeping only the terms with non-zero matrix elements in the physical subspace, i.e., at most one boson per site):
\begin{align} \label{Eq:bosonham}
   H = E_0  &+ \frac{1}{4}\sum_{i} \sum_{r=0}^1  \left(\Gamma^{xx} - i\Gamma^{xy} -i \Gamma^{yx} -\Gamma^{yy}\right)a_{i+r}b_i \nonumber \\
&\quad
    +\frac{1}{4}\sum_{i} \sum_{r=0}^1\left(\Gamma^{xx} + i\Gamma^{xy} -i \Gamma^{yx} +\Gamma^{yy}\right)b^{\dagger}_ia_{i+r} \nonumber \\
&\quad +\frac{1}{4}\sum_{i} \sum_{r=0}^1\left(\Gamma^{xx} + i\Gamma^{xy} +i \Gamma^{yx} -\Gamma^{yy}\right)a^{\dagger}_{i+r}b^{\dagger}_i \nonumber \\
&\quad+\frac{1}{4}\sum_{i} \sum_{r=0}^1\left(\Gamma^{xx} - i\Gamma^{xy} +i \Gamma^{yx} +\Gamma^{yy}\right)a^{\dagger}_{i+r}b_i \nonumber \\
&\quad + \sum_i\left(h R^{zz} - \Gamma^{zz} \right)\left(a^{\dagger}_ia_i + b^{\dagger}_i b_i \right) + H_{I},
\end{align}
where $\Gamma^{\alpha \beta} = -J\left(R^{x\alpha}_A R^{x \beta}_B + R^{y\alpha}_A R^{y \beta}_B +R^{z\alpha}_A R^{z \beta}_B\right) + D_x \left(R^{y\alpha}_A R^{z \beta}_B -R^{z\alpha}_A R^{y \beta}_B\right)$ depends on the particular local rotated frames chosen for the $A$ and $B$ sublattices, given by rotation matrices $R_{A(B)}$. Furthermore, the operators $a_i$ and $b_i$ destroy a boson on the $A$ and $B$ sublattice on site $i$, respectively. Note that the $\Gamma$ factors will change when we expand around different energy extrema (compare Sec. \ref{sec:otherExtrema}) to compensate for the fact that the definition of the bosons has changed.

$H_I$ is an interaction term involving cubic and quartic boson operators:
\begin{align*}
 &H_I = \sum_{i} \sum_{r=0}^1 \frac{i}{2} \Gamma^{yz} \left(b^{\dagger}_ib_ia_{i+r} - b^{\dagger}_i a^{\dagger}_{i+r}b_i \right) \\
 &\quad +\sum_{i} \sum_{r=0}^1\frac{i}{2}\Gamma^{zy} \left(a^{\dagger}_{i+r}a_{i+r}b_i - a^{\dagger}_{i+r}b^{\dagger}_ia_{i+r} \right) \\
 &\quad +\sum_{i}\sum_{r=0}^1\Gamma^{zz}a^{\dagger}_{i+r}b^{\dagger}_ib_ia_{i+r}.
\end{align*}
    
Significantly, linear terms in boson operators drop out as long as we choose the local frames carefully - a critical case in which this happens is when one chooses the classical ground state configuration. However, we stress that one also loses all bosonic linear terms by expanding around any extremum of the classical energy. This observation helps access states at the maximum end of the energy spectrum while retaining a single-particle or magnon language.

For illustrative purposes, we note that when we restrict to a strictly single-boson subspace (the lowest nontrivial order in our approximation scheme), the Hamiltonian takes the form 
\begin{align*} \label{eq:single}
    \tilde{H} = E_0 &
    +\frac{1}{4}\sum_{i} \sum_{r=0}^1\left(\Gamma^{xx} + i\Gamma^{xy} -i \Gamma^{yx} +\Gamma^{yy}\right)b^{\dagger}_ia_{i+r} \\
&\quad+\frac{1}{4}\sum_{i} \sum_{r=0}^1\left(\Gamma^{xx} - i\Gamma^{xy} +i \Gamma^{yx} +\Gamma^{yy}\right)a^{\dagger}_{i+r}b_i \\
&\quad + \sum_i\left(h R^{zz} - \Gamma^{zz} \right)\left(a^{\dagger}_ia_i + b^{\dagger}_i b_i \right),
\end{align*}
where $\tilde{H}$ contains only the terms 
that have non-zero single-boson matrix elements and thus contribute to $H^{(1)}$. In other words, when we restrict to the region of the Hilbert space containing at most single-boson states, normal-ordered terms containing more than one creation operator or more than one annihilation operator drop out, as these terms destroy any state in the single-boson subspace. 

Higher orders in the approximation scheme are captured by diagonalizing $H^{(n)}$, in which we consider the subspace with up to $n$ total bosons.

\section{The magnetic ground state}
\label{Sec:mag_gnd_state}
Having discussed the general form of the HP transformation that leads to the conventional linear spin-wave Hamiltonian, it is useful to pause and recall the structure of the magnetic ground state within this approximation, which is in general nontrivial to compute; we will see that our scheme provides a more succinct approximation of the magnetic ground state for an arbitrary spin Hamiltonian. 

To motivate this point, we review the computation of the ground state of the linear spin wave theory, providing helpful context for our own results presented in \ref{Sec:Results}, which are in excellent agreement with the exact result. While much of this material will be well known to experts, we include a detailed discussion to make the work self-contained.

After transforming the spin Hamiltonian into the bosonic Hamiltonian in Eq.\eqref{Eq:bosonham}, the linear spin-wave approximation proceeds by dropping the interaction term $H_{I}$. This step results in a bilinear Hamiltonian that can be diagonalized. For a translationally invariant system, it is often useful to first apply a Fourier transformation to block-diagonalize the Hamiltonian and cast it in the form 
\begin{equation}
    H = \sum_k \psi^{\dagger}_k H_k \psi_k.
\end{equation}

For a system with $N$ internal degrees of freedom (such as sublattice, orbital, or spin), the Nambu spinors $\psi_k$ are given as
$$\psi_{k} = \begin{pmatrix}
     a_1(k) \\
     \vdots \\
        a_N(k) \\
        a^{\dagger}_1(-k) \\
        \vdots \\
        a^{\dagger}_N(-k) 
\end{pmatrix}.$$

One may now diagonalize $H_k$ to solve the problem, such that the Hamiltonian in its diagonal form is given as
\begin{equation}
    H = \sum_{k} \sum_{n = 1}^N E_n(k) \gamma^{\dagger}_n(k) \gamma_n(k),
\end{equation}
where this diagonalization transformed the HP bosons (in the Fourier representation) $a_n(k)$ into Bogoliubov bosons (magnons) $\gamma_n(k)$ with corresponding excitation energies $E_n(k)$. To ensure that the operators $\gamma_n(k)$ still obey bosonic commutation relations, this diagonalization must be performed via a paraunitary transformation. This requirement originates from the fact that the Hamiltonian contains pairing terms such as $a_n(k)a_m(k)$ which do not conserve particle number \footnote{This observation is different from the case of fermions where a unitary transformation conserves the fermionic algebra. This difference is tied to the observation that bosons obey commutation rather than anti-commutation relations. That is, the algebra is not conserved under a transformation $a_i^\dagger \leftrightarrow a_i(-k)$. As a result, unlike for fermions, creation operators cannot be reinterpreted as annihilation operators. This fact ultimately leads to the observation that commutation relations after a unitary transformation that mixes creation and annihilation operators fail in the bosonic case. It is essentially due to the negative sign from switching the order of $a$ and $a^\dag$ in the commutators.}. 

We can write the paraunitary transformation as 
\begin{equation} \label{Eq:transformation}
    \vec{\gamma}(k) = u(k) \vec{a}(k) + v(k) \vec{a}^{\dagger}(-k),
\end{equation}
where $u$ and $v$ are $N \times N$ matrices and $$\vec{\gamma}(k) = \begin{pmatrix}
     \gamma_1(k) \\
     \vdots \\
        \gamma_N(k)
\end{pmatrix}.$$
It is well known that this transformation does not preserve the number operator, so $\sum_n a^{\dagger}_n a_n \neq \sum_n \gamma^{\dagger}_n \gamma_n$.

With all this in place, we now investigate the system's ground state from the perspective of the linear spin-wave approximation. When expanding around the classical ground state, we expect the quantum ground state to be near (but not identical to) this magnetically ordered state. In the HP representation, this is the state with no HP bosons - the $z$-component of the spin points along the classical configuration. However, when we diagonalize to obtain Bogoliubov bosons, our ground state is not the HP boson vacuum but the Bogoliubov vacuum. Since the boson number is not preserved across the transformation, it is understood that these two vacua will not generally be identical; for example, the ground state of an antiferromagnet contains quantum fluctuations around the ordered N\'eel state. 

To make this statement more formal, we can begin with the form in Eq.\eqref{Eq:transformation}. We seek to represent the Bogoliubov vacuum, such that $\gamma_n(k) \ket{\Omega} = 0$, in terms of the HP vacuum $a_n(k)\ket{0} = 0$. Though not always globally true, for many $k$, $u(k)$ will be invertible. At those points, we can write (suppressing the $k$-dependence for now and summing over repeated indices)
\begin{equation} \label{Eq:condition}
    0 = \gamma_n \ket{\Omega} = u^{-1}_{mn}\gamma_n \ket{\Omega}
\end{equation}
$$= (a_m - T_{mn}a^{\dagger}_n)\ket{\Omega},$$
where $T = -u^{-1}v$. Next, we define an operator $Q = \frac{1}{2} \sum_{ij}T_{ij}a^{\dagger}_i a^{\dagger}_j$ \cite{StoneBogoliubov}. This operator has the property that the commutator $[Q, a_i] = -T_{ij}a^{\dagger}_j$ itself commutes with $Q$. This observation implies that an expansion of $e^{Q}a_ie^{-Q}=e^{[Q,.]}a_i$ in terms of nested commutators terminates\footnote{Here we used properties of the adjoint map to write it in terms of an exponential of nested commutators. Sometimes, in the literature, $[Q,.]$ would be written as $ad_{Q}$. This notation, in particular, means $[Q,.]^2 A=[Q,[Q,A]]$ etc.}:
$$e^{Q}a_ie^{-Q} = a_i + [Q, a_i] = a_i -T_{ij}a^{\dagger}_j.$$

We find that Eq.\eqref{Eq:condition} is satisfied if and only if $e^Q \ket{\Omega} \propto \ket{0}$. Because this must hold for each $k$ (certainly for the points at which $u$ is invertible), we find finally that:
\begin{equation} \label{Eq:GS}
    \ket{\Omega} = \mathcal{N} \prod_k e^{-Q(k)} \ket{0}.
\end{equation}

This gives us an explicit form for the Bogoliubov vacuum in terms of HP boson operators and vacuum. As expected, Eq.\eqref{Eq:GS} shows that the Bogoliubov vacuum is a complicated superposition of states with indefinite HP particle number. Indeed, because $Q$ can be quite difficult to compute in closed form, finding this approximate ground state can involve significant analytical overhead, and one must take care not to include non-physical occupation numbers. By contrast, our method provides an analytically straightforward means of approximating the spin system's ground state, both efficiently and with high agreement with numerically exact results obtained in the 1D case. 

We also note that due to the inclusion of unphysical states in $\ket{\Omega}$, this state will necessarily involve some degree of error that is not shared with our approximation for the ground state, which does not involve unphysical contributions. For a system of spins with spin $S$, the maximum boson occupation per site is $2S$. However, the Bogoliubov calculation involves no bounds on occupation. Thus, the result Eq.\eqref{Eq:GS} will involve leakage into the unphysical part of the boson Fock space; in other words, the amplitudes of the physical states that contribute to the exact solution will necessarily be smaller in the Bogoliubov approximation. 

\section{Validity of Hilbert space truncation approximation} \label{Sec:Controlled}
Before presenting the results of the Hilbert space approximation method, we turn to the question of whether the approximation is well controlled. In other words, for small canting of the classical ground state configuration (which takes us away from an exactly solvable Hamiltonian), does the contribution from different occupation sectors increase in a controlled way? 

Our explicit spin chain model helps answer this question because the magnitude of the canting is directly related to the size of the DMI; we may, therefore, tune the size of the number of non-conserving terms through the DMI magnitude. In other words, the model permits us to gain controlled insights into the validity of our approximation scheme.

We note that in our case, the approach of working in a Hilbert space that is truncated at a maximum number of bosons is very similar to quasi-degenerate perturbation theory, where one considers not just a single state of interest but rather expands the Hamiltonian for a subset of states that can perturbatively be expected to be most strongly coupled to the state of interest and diagonalizing exactly in this subspace. In other words, we diagonalize the Hamiltonian within the (nearly) degenerate subspace. 
In our case, perturbations at the lowest orders (as we will see later) lead to couplings with new particle sectors, which we, in the spirit of a quasi-degenerate perturbation theory, include. The advantage of such a scheme is that it (as we will see later) avoids singularities that may appear due to possibly existing degeneracies. Moreover, suppose the space that one truncates for is large enough. In that case, one may obtain results beyond the reaches of perturbation theory - diagonalizing the truncated Hamiltonian corrections to energy and states are not merely polynomials of the perturbation parameter. In this sense, one may alternatively think of this approach as a partial resummation, where parts of the Hilbert space were treated exactly. 

Furthermore, one may recall that quasi-degenerate perturbation theory leads to results with ordinary perturbation theory as a limit in the case of non-degenerate energy states. Therefore, to get an estimate of the validity of our approach (a lower bound on its validity), we begin by analyzing contributions from disparate number sectors within the Brillouin-Wigner perturbation theory. 

In what follows, we primarily perform our approximation scheme with respect to the HP representation. However, the process proceeds identically within the Bogoliubov (magnon) representation, obtained by performing the paraunitary transformation which diagonalizes the quadratic Hamiltonian on the rest of the terms that appear in the fully interacting bosonic Hamiltonian. 

\subsection{Brillouin-Wigner perturbation theory}
As a first step, we review Brillouin-Wigner perturbation theory \cite{refId0, Wigner1997}. Like the more conventional Rayleigh-Schr\"odinger perturbation theory \cite{Sakurai}, it is articulated with respect to an unperturbed problem
\begin{equation}
    H_0\ket{n} = \epsilon_n \ket{n}.
\end{equation}
This Hamiltonian has a spectral representation $H_0 = \sum_n \epsilon_n \ket{n}\bra{n}$, which is diagonal. We can define a projection operator 
\begin{equation}
    Q_n = I - \ket{n}\bra{n} = \sum_{m\neq n} \ket{m}\bra{m},
\end{equation}
which commutes with $H_0$ and is, therefore, a conserved quantity in the unperturbed theory. Furthermore, $Q_n$ projects onto the subspace of states that differ from $\ket{n}\bra{n}$.

The perturbed eigenvalue problem for $H = H_0 + H_1$ has the form:
\begin{equation}
    H\ket{\psi_n} = E_n \ket{\psi_n},
\end{equation}
which can be rewritten as 
\begin{equation}
    (E_n - H_0)\ket{\psi_n} = H_1\ket{\psi_n}.
\end{equation}
Acting on the left with $Q_n$, we find
$$Q_nH_1\ket{\psi_n} = (E_n - H_0)Q_n\ket{\psi_n},$$
\begin{equation}
    (E_n-H_0)^{-1} Q_n H_1 \ket{\psi_n} = Q_n \ket{\psi_n}.
\end{equation}
We then define 
\begin{equation}
    R_n = (E_n-H_0)^{-1} Q_n = \sum_{m\neq n} \frac{\ket{m}\bra{m}}{E_n - \epsilon_m},
\end{equation}
where the last equal sign is obtained by switching to a spectral representation. We then have
\begin{equation}
    R_n H_1 \ket{\psi_n} = Q_n \ket{\psi_n}.
\end{equation}
Writing the perturbed state out in terms of unperturbed states via the insertion of an identity, we have
\begin{equation}
    \ket{\psi_n} = \sum_m \ket{m} \bra{m} \ket{\psi_n} = \ket{n}\bra{n}\ket{\psi_n} + Q_n \ket{\psi_n}
\end{equation}
$$=\ket{n}\bra{n}\ket{\psi_n} + R_n H_1\ket{\psi_n}.$$
If we use the convention $\bra{n}\ket{\psi_n} = 1$, thereby working with unnormalized $\ket{\psi_n}$, then we have
\begin{equation}
    \ket{\psi_n} = \ket{n} + R_n H_1 \ket{\psi_n},
\end{equation}
which we can solve iteratively by plugging the right side into itself. This idea should lead to a sensible perturbative scheme if $H_1$ is small relative to $H_0$. We obtain
\begin{equation}
     \ket{\psi_n} = \ket{n} + R_n H_1 \ket{n} + (R_n H_1)^2 \ket{n} + \dots
\end{equation}

For our purposes, we wish to show that contributions from different number sectors of Hilbert space to a given state are perturbatively small in the small canting limit, i.e., $D \to 0$. It is, therefore, useful to explicitly include labels for particle number $N$ such that we find
\begin{equation} \label{eq:pert}
    \ket{\psi_{N,a}} = \ket{N,a} + R_{N,a} H_1 \ket{N,a} + \dots
\end{equation}
at lowest order, where $R_{N,a} = \sum_{\{N,a\} \neq \{M,b\}} \frac{\ket{M,b}\bra{M,b}}{E_{N,a} - \epsilon_{M,b}}$, $\ket{N,a}$ labels a state with $N$ total bosons, and $a$ is a quantum number or set of quantum numbers that express the degrees of freedom within the $N$-boson sector. $E_{N,a}$ is the exact energy of the perturbed state being approximated. At the same time, $\epsilon_{M,b}$ is an unperturbed energy - we do not need to know either of these energies.

\subsection{Controlled expansion}
Now, Eq.\eqref{eq:pert} is the basis for arguing that the Hilbert space truncation approximation we advocated for earlier is well-behaved. One way to assess this is to take the boson-transformed Hamiltonian and arrange it according to terms that are number-conserving and those that are not. The number-conserving terms are much larger for small canting (small DMI). This observation allows for a perturbative justification of the expansion, assuming there are not so many number non-conserving terms that, taken together, they are not negligible compared to the number-conserving terms. 

To see this, we refer back to Eq.\eqref{Eq:bosonham} and collect all of the number non-conserving terms into $H_1$ so that
\begin{dmath} \label{eq:H1abstract}
    H_1 =  \frac{1}{4}\sum_{i} \sum_{r=0}^1  \left(\Gamma^{xx} - i\Gamma^{xy} -i \Gamma^{yx} -\Gamma^{yy}\right)a_{i+r}b_i + \frac{1}{4}\sum_{i} \sum_{r=0}^1\left(\Gamma^{xx} + i\Gamma^{xy} +i \Gamma^{yx} -\Gamma^{yy}\right)a^{\dagger}_{i+r}b^{\dagger}_i+\sum_{i} \sum_{r=0}^1 \frac{i}{2} \Gamma^{yz} \left(b^{\dagger}_ib_ia_{i+r} - b^{\dagger}_i a^{\dagger}_{i+r}b_i \right) +\sum_{i} \sum_{r=0}^1\frac{i}{2}\Gamma^{zy} \left(a^{\dagger}_{i+r}a_{i+r}b_i - a^{\dagger}_{i+r}b^{\dagger}_ia_{i+r} \right).
\end{dmath}

$H_0$, therefore, contains only number-conserving terms, which justifies taking our unperturbed states to have definite particle numbers. In this case, we can write to first order:
\begin{equation}
     \ket{\psi_{N,a}} = \ket{N,a} + \sum_{\{N,a\} \neq \{M,b\}} \frac{\bra{M,b} H_1 \ket{N,a}}{E_{N,a} - \epsilon_{M,b}} \ket{M,b},
\end{equation}
where we see that the matrix elements $\bra{M,b} H_1 \ket{N,a}$ determine the size of the $\ket{M,b}$ component of the corrected state. We can, therefore, easily look for matrix elements such as $\bra{2,b} H_1 \ket{1,a}$ to tell us (for example) the size of a contribution from two-particle space to an unperturbed single-particle state.

These elements are straightforward to read off from $H_1$. The first and second terms couple sectors that are two particles apart, while the third and fourth couple sectors are one particle apart. For this model, no other sectors are coupled. We find in particular that a state $\ket{N,a}$ receives contributions of order $\Gamma^{zy}$ from sectors one particle away and of order $\Gamma^{xx} - \Gamma^{yy}$ from sectors two particles away. 

If we consider the expansion around the classical ground state, we can write down the dependence of these terms on the DMI strength $D$ through the canting angle $\phi$. To do this, we first assume that the canting angle is small, as it is for a small DMI. In this case, we find that the energy is minimized by 
\begin{equation}
    \phi \approx \frac{-(J + \Delta J + h) + \sqrt{(J + \Delta J + h)^2 + 8 D^2}}{4D}.
\end{equation}
Further expanding the square root for small $D$, we have
\begin{equation}
    \phi \approx \frac{D}{J+\Delta J +h}.
\end{equation}
Therefore, we can write down an expression for $H_1$ in terms of the spin coupling parameters, valid for small $D/J$. Letting $\tilde{J} = J + \Delta J + h$, we can write
\begin{align*} \label{eq:H1abstract}
    H_1 =  &\frac{1}{4}\sum_{i} \sum_{r=0}^1  \left( -2J\frac{D^2}{\tilde{J}^2} -2\frac{D^2}{\tilde{J}}\right)(a_{i+r}b_i + a^{\dagger}_{i+r}b^{\dagger}_i)\\
    +&\sum_{i} \sum_{r=0}^1 \frac{i}{2} \left( 2J\frac{D}{\tilde{J}} + D -2\frac{D^3}{\tilde{J}^2}\right) \left(b^{\dagger}_ib_ia_{i+r} + a^{\dagger}_{i+r}b^{\dagger}_ia_{i+r}\right) \\ +& {\rm{h.c.}},
\end{align*}
where h.c. denotes the Hermitian conjugate of the second line. This expression - for the case of small $D/J$ - makes it clear that the contributions from disparate number sectors will be perturbatively small in the limit $D \to 0$. However, it also informs us of the relative importance of different number sectors in our truncation approximation. We find that contributions from sectors one occupation number away are the strongest, of order $D$, while contributions from two are of order $\frac{D^2}{J}$. Interestingly, to first order in perturbation theory, there are no contributions from particle sectors more than two occupation numbers away. 

We can also compute higher orders in perturbation theory, where multiple ``steps" between particle sectors are now allowed. The second-order contribution is, for example, given as
\begin{widetext}
\begin{equation}
    \sum_{M',b'\neq N',a'}\sum_{M,b \neq N,a} \ket{M,b} \frac{1}{(E_{N,a} - \epsilon_{M,b})(E_{N,a} - \epsilon_{M',b'})} \bra{M,b} H_1 \ket{M',b'} \bra{M',b'} H_1 \ket{N,a}.
\end{equation}
\end{widetext}
To summarize, truncating the Hilbert space for weak perturbations is justified because only states from a few particle sectors contribute to wavefunctions. This observation justifies using quasi-degenerate perturbation theory to truncate our Hilbert space.

\subsection{Degeneracies in the Brillouin-Wigner series}
We stress that, strictly speaking, the argument above applies in cases where the energies $\epsilon_{M,b}$ do not have degeneracies between number sectors. Note that no problems arise for degeneracies {\ within} a particle sector because $H_1$ has no non-zero matrix elements within an individual sector - it only couples states in different sectors. 

If $\epsilon_{N,a} = \epsilon_{M,b}$ for some $M \neq N$, then $\lim_{D \to 0} (E_{N,a} - \epsilon_{M,b}) = 0$. In this case, the standard procedure of degenerate perturbation theory is to work in a basis in which $H_1$ is diagonal within the degenerate subspace. However, this is what our approximation procedure already does - we diagonalize within a subspace. As long as the degenerate states are within the subspace, we are working in, the argument above proceeds safely. Therefore, one does not have to worry about energy degeneracies for our procedure.

\section{Results} \label{Sec:Results}

\subsection{Efficacy of approximation scheme}
\begin{figure}[htp]
\centering
    \includegraphics[width=8cm]{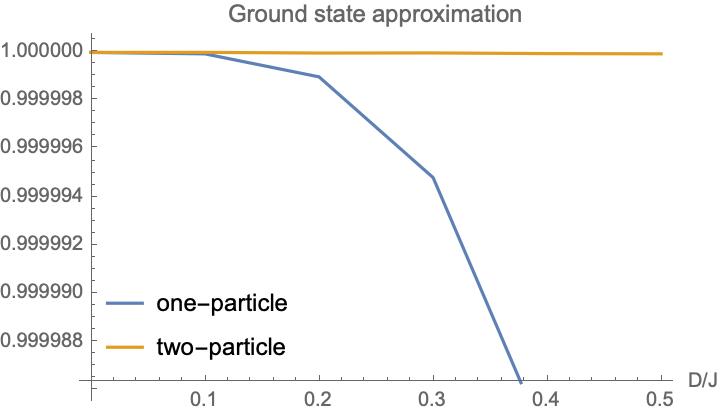}
    \includegraphics[width=8cm]{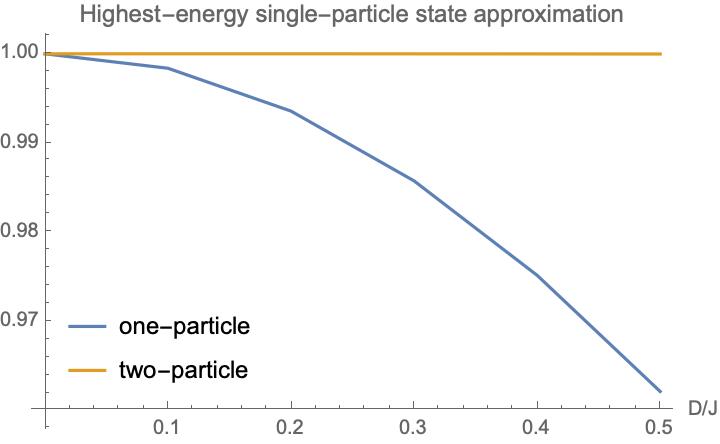}
    \includegraphics[width=8cm]{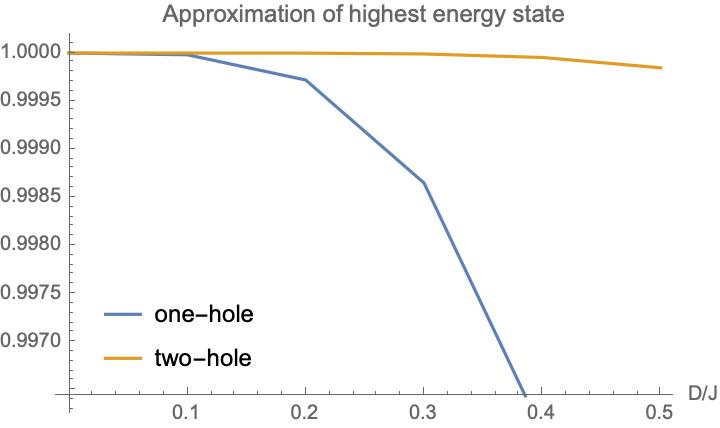} 
   \caption{\label{Fig:Overlaps6site}Approximation of eigenstates in one- and two-particle truncated Hilbert spaces for a 6-site model. Plotted on the $y$-axis is the overlap of the approximate state with the exact one (computed by exact diagonalization). } 
    \end{figure}

Here, we present results comparing our truncated Hilbert space approximation scheme to exact diagonalization results. Overlaps between an exact and approximate wavefunction at various energies are shown for a six-site model in Fig.\ref{Fig:Overlaps6site}. We chose to consider overlaps between wavefunctions because they are a more reliable measure of approximation goodness than closeness in energies (e.g., variational energies can be quite accurate even when the variational wavefunctions are not). As anticipated in Sec.\ref{Sec:Controlled}, we find that truncating at two bosons already produces nearly exact solutions, owing to the structure of the interaction terms in the Hamiltonian of  Eq.\eqref{Eq:bosonham}, which link at most two particle sectors. 

Furthermore, the approximation is slightly worse at higher energy scales and for larger DMI strength. However, This observation is expected and arises from the fact that for larger $D/J$, the system is farther from the exactly solvable, number-conserving case. 
It is worth noting that the approximation does well even at $D/J = 0.5$, especially in the two-particle and two-hole sectors. In real spin systems, $D$ tends to be smaller than $J$, owing to its origin in spin-orbit coupling \cite{PhysRev.79.350, PhysRev.120.91, Mazurenko2021-yc}. 

\begin{figure}[htp]
\centering
    \includegraphics[width=8cm]{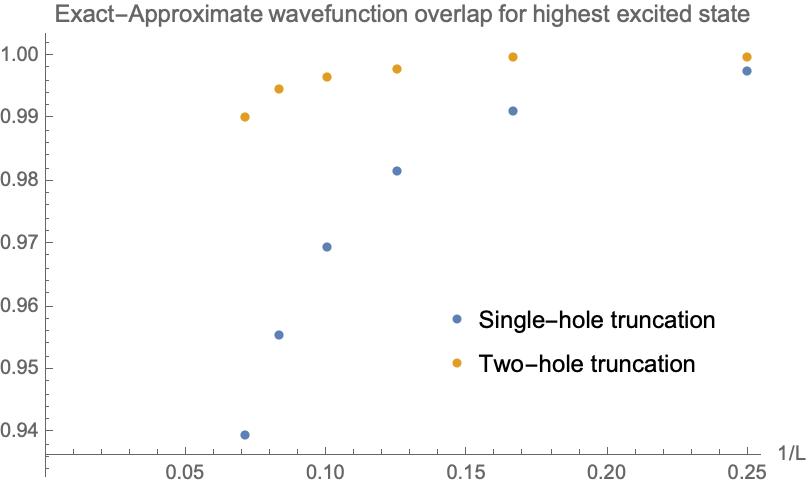}
   \caption{\label{Fig:length}Overlaps between exact and approximate eigenstates, computed for the highest energy state in the system, plotted against inverse system size. As $1/L \to 0$, the overlap extrapolates to cross the $y$-axis at a finite value. 
   } 
    \end{figure}

\begin{figure}[htp]
\centering
    \includegraphics[width=8cm]{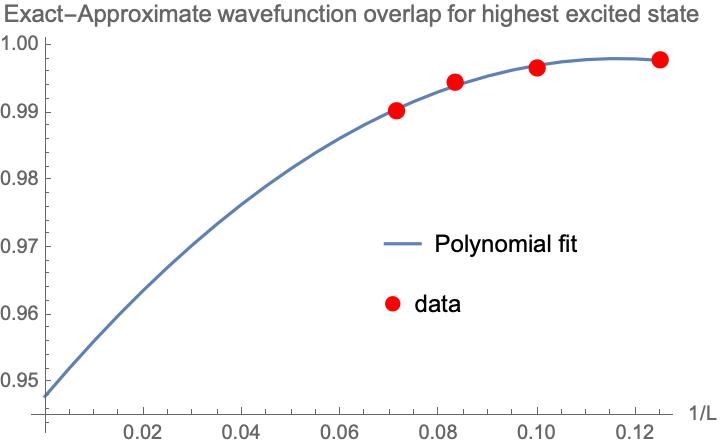}  
   \caption{\label{Fig:sizefit} To verify that the overlap between exact and approximate wavefunctions is finite in the thermodynamic limit, we fit the data for the largest system sizes to a polynomial. The fit crosses the $y$-axis ($1/L \to 0$) at a finite value, $\sim 0.948$. We note that when the same fit function is used for only the three (rather than four) largest system sizes, we find a $y$-intercept $\sim 0.917$. This supports the conclusion that in the thermodynamic limit there is a finite $y$-intercept. This crossing point will, of course, have some dependence on the fit function, but here, we are more concerned with the existence of a non-zero $y$-intercept than with its particular value. We wish to see that in the $L\to \infty$ limit, the approximation has a finite overlap with the exact result even for a low-order approximation involving just up to two bosons.} 
    \end{figure}

It can reasonably be asked whether the significant agreement with exact results is a feature of small system size rather than a robust approximation scheme. We check this by comparing results across system sizes from 6 to 12 sites, as shown in Fig.\ref{Fig:length}. We plot the overlap against inverse system size $1/L$ to extrapolate to $1/L=0$ . 
The overlap between the exact and approximate results appears to converge to a nonzero value (again, especially for results obtained in the two-hole subspace). This is quantitatively supported by fitting the data for the largest system sizes to a second-order polynomial to extrapolate to the thermodynamic limit, as shown in Fig.\ref{Fig:sizefit}.


\begin{figure}[htp]
\centering
    \includegraphics[width=8cm]{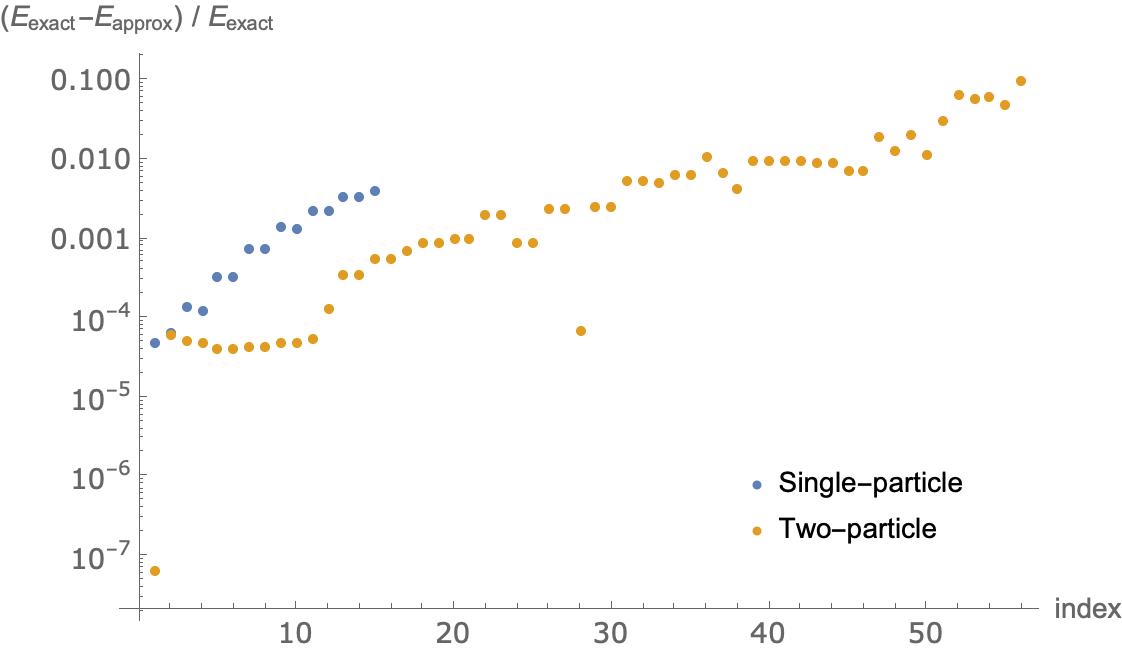}   
   \caption{\label{Fig:energy} Relative error in approximating energy eigenvalues for a 14-site model, for a single-particle truncation and a two-particle truncation. The $y$-axis is plotted on a log scale. The index on the $x$-axis is a quantum number labeling distinct energy eigenstates.} 
    \end{figure}

Finally, in addition to eigenstates, obtaining approximate eigenenergies of the system is often helpful. We find again that the approximation is best at lower energies within a given subspace and that the two-particle truncation substantially improves upon the single-particle one, as seen in Fig.\ref{Fig:energy}. 

\subsection{Numerical savings}
In addition to the efficacy of the approximation scheme in various regimes, an important practical element of any approximation computed numerically is the computational savings it allows compared to an exact solution. We therefore use full exact diagonalisation studies as our yardstick, whose scaling behaviour is well known independently of dimension. Of course, in low dimension, especially 1D, there exist more sophisticated methods with considerable speedup. However, these do not readily generalize to higher dimension or higher-energy starting states, while the generalization of the method we present here is straightforward. In general, the time to diagonalize an $N$-dimensional matrix asymptotically is $ \mathcal{O}(N^3)$. A system with $L$ sites will have a Hilbert space of size $(2S+1)^L$, leading to computing times of $\sim \mathcal{O}((2S+1)^{3L})$. Thus, increasing the system size leads to an exponential increase in the computing time. 

Selective Hilbert space truncation, therefore, leads to significant numerical savings. For example, restriction to globally $\leq1$ boson amounts to solving a problem with dimension $L+1$ instead of dimension $(2S+1)^L$, which drastically reduces numerical overhead. The corresponding computing time is then $\sim \mathcal{O}(L^{3})$, with cubic rather than exponential scaling in system size. As discussed in the previous section, this speedup is achieved while maintaining proximity to exact results.

Truncating to $\leq2$ bosons globally also dramatically reduces the size of the diagonalization problem while significantly increasing the approximation's efficacy, as discussed above. In this case, the size of the truncated problem for a spin-1/2 system is $(L+1) + \frac{1}{2} \frac{L!}{(L-2)!}$, and for spin $S>1/2$ it is $2(L+1) + \frac{1}{2} \frac{L!}{(L-2)!}$. Therefore, the system size grows asymptotically as $L^2$, for a computation time of $\mathcal{O}(L^6)$. Thus, the reduced problem has polynomial scaling with system size, while the full one has exponential scaling. This property is especially relevant for the computation of magnon-bound states and highlights that this approximation method allows for significant numerical savings in the computation of these states.

\begin{figure}[htp]
\centering
    \includegraphics[width=8cm]{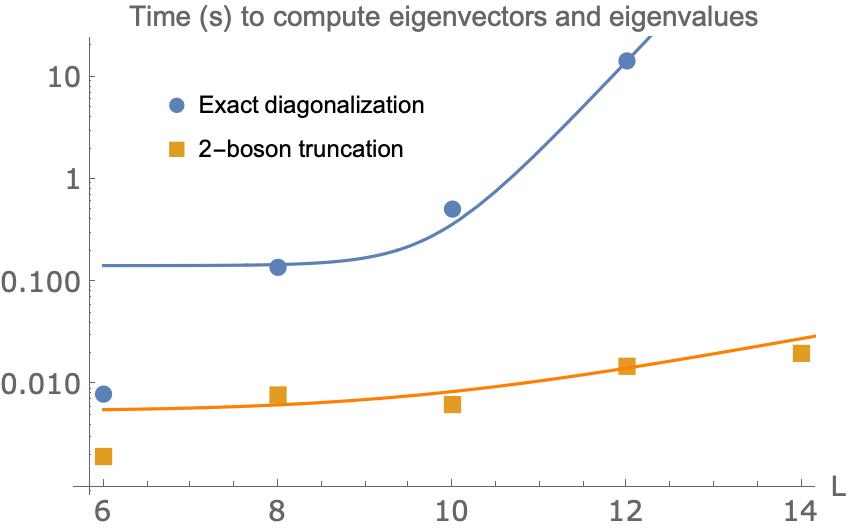}
   \caption{\label{Fig:time} Run time to compute eigenvalues and eigenstates via exact diagonalization for exact and truncated Hilbert spaces for up to two bosons on a log scale. We see a drastic speedup in the truncated problem. Best fit curves for the asymptotic behavior are also shown, $\mathcal{O}(2^{3L})$ for the exact calculation and $\mathcal{O}(L^6)$ for the approximation.} 
    \end{figure}

Fig.\ref{Fig:time} allows for a concrete visualization of this effect, comparing the exact diagonalization performed for an entire system and the run time to compute the truncated eigenspectrum. The truncation is done for $\leq2$ bosons globally for a spin-1/2 system. 

\subsection{Application to boson bound states}

Finally, we present an important application of the Hilbert space truncation scheme - the study of boson-boson bound states. We first present a discussion of Holstein-Primakoff boson bound states, before commenting on our method's applicability to active research around magnon bound states \cite{Fukuhara2013-yl,PhysRevLett.95.077201,PhysRevLett.100.027206,PhysRevLett.102.037203,PhysRevB.67.054414,PhysRevB.108.064408,PhysRevLett.127.267201,PhysRevLett.124.197203,Nishida2013-zc,PhysRevResearch.2.033024,Wang2018-yy,PhysRevLett.125.187201,PhysRevB.101.180401,Wulferding2020-jf,PhysRev.132.85,H_C_Fogedby1980-xe}. Research on magnon bound states has two primary difficulties. Experimentally, systems with magnetic order often have magnon-bound states with lower energy than their single-magnon counterparts; this can lead to difficulty probing these states effectively \cite{PhysRevB.67.054414}. Theoretically, there is a need for techniques to characterize and efficiently compute magnon bound state properties.

The scheme presented here provides solutions to each of these problems. First, the transformation to an exact bosonic Hamiltonian (either in HP bosons or magnons themselves) allows us to identify phases of a given model in which low-lying bound states will exist - in particular, bound states that are lower in energy than their single-boson counterparts. Second, our approximation scheme allows for the efficient characterization of the bound states themselves.

Identifying phases with low-lying bound states opens the door to theoretical proposals for systems with this property. This identification can be done at the level of the exact bosonic Hamiltonian - in particular by looking for phases in which interaction terms lead to energy savings for pairs of bosons over individual ones. In what follows, we analyze bound states in the HP representation. Analysis of magnon bound states proceeds identically after performing a Bogoliubov transformation on the $k$-space HP Hamiltonian. 

Taking the 1D model discussed in this work as an example, we can look in particular at the boson mass term $\left(h R^{zz} - \Gamma^{zz} \right)\left(a^{\dagger}_ia_i + b^{\dagger}_i b_i \right)$ and the interaction term $\Gamma^{zz}a^{\dagger}_{i+r}b^{\dagger}_ib_ia_{i+r}$. In order to have boson bound states, the interaction must be attractive so that $\Gamma^{zz} <0$. At the same time, for the bound states to be low-lying, the energetic cost of adding a particle must be outweighed by the reduction due to boson attraction. Thus, the mass coefficient  $\left(h R^{zz} - \Gamma^{zz} \right)$ must be small. When these conditions are met, the spectrum exhibits low-lying bound states with lower energy than at least some of the system's single-boson states. Moreover, we see that for this model, the applied field $h$ can be used to drive bound states lower in the spectrum. Similar phenomena are likely to be found in higher-dimensional systems and real materials, allowing for experimental creation and detection of these bound states more readily.

While such an analysis could have been done on a bosonic Hamiltonian obtained in the conventional way using a truncated Taylor series in the HP bosons, significant error is introduced in the coefficients of the interaction terms, especially for small spin (see Tab. \ref{Tab:interaction}). A Taylor series expansion gives

\begin{equation} \label{TaylorHP}
    S^{+} = \sqrt{2S} (1 - \frac{a^{\dagger}a}{2S})^{1/2}a
\end{equation}
$$= \sqrt{2S}(a - \frac{1}{4S}a^{\dagger}a^2 + \mathcal{O}(1/S^2)).$$

By contrast, the number of terms in the exact expansion is finite but depends on the value of $S$. For $S = 1$, we have (keeping only terms that act nontrivially within the physical Hilbert space):

\begin{equation}
    S^+ = \sqrt{2}a - (\sqrt{2}-1)a^{\dagger}a^2.
\end{equation}
Writing $S^+ = \sqrt{2S}a - \lambda a^{\dagger}a^2$, we note that $\lambda$ determines the strength of interaction terms in a spin Hamiltonian with the form $H = \sum_{ij} \Gamma^{\alpha \beta}S^{\alpha}_i S^{\beta}_j$, aside from the simple attraction/repulsion term $\Gamma^{zz}a^{\dagger}_{i+r}b^{\dagger}_ib_ia_{i+r}$ discussed above. 

The error introduced by the conventional Taylor series truncation is seen clearly for $S=1$, in which case $\lambda = \sqrt{2} - 1$, while the truncated Taylor series gives $\lambda = \sqrt{2}/4$. We illustrate the trend in this discrepancy as we go to larger spin in Table \ref{Tab:interaction}.

\begin{table}
\begin{tabular}{|c|c|c|c|c|c|c|}
\hline

   $S$ & $1/2$ & $1$ & $3/2$ & $2$ & $5/2$ & $3$\\ \hline $\lambda$  & 1.0 & 0.41 & 0.32 & 0.27 & 0.23 & 0.21\\ \hline
    $\lambda_T$ & 0.50 & 0.35 & 0.29 & 0.25 & 0.22 & 0.20\\ \hline 
\end{tabular}

\caption{Exact values of $\lambda$ compared to values $\lambda_T$ obtained by a truncated Taylor series, the conventional method for treating problems in which boson interactions are important. We see that the truncated Taylor series underestimates the size of $\lambda$, thereby overestimating the stability of multi-boson bound states. \label{Tab:interaction}}
\end{table}

 Interaction terms involving factors of $\lambda$ encode the dynamics of multi-boson states, including propagation and decay/production. For example, we find terms with the form $a^{\dagger}_j a^{\dagger}_i a^2_j$, allowing two bosons occupying the same site to evolve into a state in which the two bosons occupy neighboring sites $i$ and $j$. We also find decay terms in which a two-boson state evolves into a single-boson state and vice versa - these terms clearly will impact the stability of multi-boson bound states. In particular, these terms are incorrectly scaled in a conventional Taylor series. For $S = 1$, the Taylor series will give quartic terms with $\approx 17 \%$ error, and terms involving six creation/annihilation operators ($\propto \lambda^2$) will have $\approx 34 \%$ error. The exact boson method thus allows for the careful and precise characterization of phases with bound states, which could have been roughly estimated using the conventional approach at best. This discussion should make it clear that if one were to investigate bound state physics using a truncated Taylor series, one would introduce significant error into computations for bound state stability and dynamics relative to the exact expansion. 

Notably, the conventional truncated Taylor series is used to study both the physics of magnon bound states, as well as any physics for which magnon interactions become important \cite{Cherepanov1993-dd,PhysRevB.109.094440,Koyama2023-ud,shi2023interacting}, such as calculations of magnon lifetimes \cite{alma9951968736501401}. The convention is to keep terms of $\mathcal{O}(\lambda)$, in which case one will have a shift in the magnon lifetime (a matrix element effect in a Fermi Golden Rule calculation) with additional lifetime modifications coming from the change in the bound state energy of the magnon as well (a density of states effect). In our approach, one begins with the finite HP boson Hamiltonian, performs a Bogoliubov transformation on all terms (including interactions), and then conducts the above analysis on single-magnon kinetic terms and magnon-magnon interactions, but without the limitations imposed by a truncated Taylor series. 

We conclude this section with an example in which a boson bound state is approximated and compared to the exact result. Within the magnetically ordered phase studied above, the mass term will decrease as the magnetic field decreases, generating bound states on the same energy scale as single bosons. This phenomenon occurs, for example, for the following parameter choice: $J = 1$, $\Delta = 0.5$, $h = 0.5$, $D = 0.5$, where in particular, we have chosen a magnetic field that is weaker than the Heisenberg coupling $J$, which sets the strength of boson attraction. For this parameter choice and an 8-site model, the second excited state is a two-boson state, with energy $-3.03$. The minimal ($\leq 2$) truncated approximation gives an eigenvalue for this state of $-3.02$, with an exact-approximate wavefunction overlap $\bra{\psi_{\rm approx}}\ket{\psi_{\rm exact}} = 0.947$, highlighting the usefulness of this approach for both identifying phases containing low-lying bound states and for computing them efficiently. By comparison, the quadratic Holstein-Primakoff approach approximates the same state with $\bra{\psi_{\rm approx}}\ket{\psi_{\rm exact}} = 0.474$. The Hilbert space truncation approach thus leads to significant improvement in the approximation of this state. 

The two approaches also lead to  different predictions about the energy ordering of these states, with the HP calculation predicting that the above two-boson state is the system's true ground state. Essentially, the limitations of the HP calculation arise due to a failure to screen off unphysical boson occupation states corresponding to spin lengths greater than $S = 1/2$. We note also that in cases where an exact solution is not available for comparison, the HP picture does not give us confidence that it should accurately capture multi-boson states, because it is agnostic to the effects of interactions. In models where interactions play a larger role, these two methods will differ drastically, with the HP calculation in general being less accurate. It is important to note as well that the exact bosonic Hamiltonian was essential in reliably finding this low-lying multi-boson state regime.

We note also that for this parameter choice, the fourth excited state is a 4-boson state, showing the method's promise for characterizing $n$-particle states.

\section{Conclusion}

We have presented a systematic approximation scheme for computing eigenstates and eigenvalues of magnetic systems at higher energies than is possible through a conventional linear spin-wave analysis. This approximation is based upon the existence of an exact representation of spin operators in terms of bosons with a finite number of terms. Our approximation is applied to a 1D spin model - an XXZ chain with DMI. 

While the analysis of a 1D system serves as a valuable proof-of-concept for the approximation scheme, the approximation itself opens avenues to studying systems that are not exactly solvable, such as higher dimensional (2D and 3D) spin systems. For example, our approach could be leveraged to define and compute quantum geometric and topological quantities for highly excited eigenstates. This task is exceptionally straightforward if one expands around higher-energy classical extrema than the classical ground state, importing results from a linear spin-wave analysis to higher-energy single-boson states. Furthermore, with an approximation of high-energy eigenvalues, it becomes possible to approximate many thermal quantities, deriving them from an approximate (truncated) partition function. Moreover, our work opens the door to efficiently studying bound magnon states.%, decreasing the computational cost from $ \mathcal{O}((2S+1)^{3L})$ to $\mathcal{O}(L^6)$, where $L$ is the number of sites in the system.

\section{Acknowledgements}
We acknowledge support from
NSF DMR-2114825 and the Alexander von Humboldt Foundation.  W.R. acknowledges support from the US Department of Defense through the NDSEG Fellowship. M.V. gratefully acknowledges the support provided by the Deanship of Research Oversight and Coordination (DROC) and the Interdisciplinery Reasearch Center(IRC) for Intelligent Secure Systems (ISS) at King Fahd University of Petroleum \& Minerals (KFUPM) for funding his contribution to this work through internal research grant No. INSS2507. This work was in part supported by the Deutsche Forschungsgemeinschaft under grants SFB 1143 (project-id 247310070) and the cluster of excellence ct.qmat (EXC 2147, project-id 390858490).

\bibliographystyle{unsrt}
\bibliography{literature}

\end{document}